\documentclass{amsart}
\usepackage{amsfonts}
\usepackage{blkarray}
\usepackage{float}
\usepackage{multirow}

\newtheorem{theorem}{Theorem}[section]

\newtheorem{proposition}[theorem]{Proposition}
\newtheorem{corollary}[theorem]{Corollary}
\theoremstyle{definition}

\theoremstyle{remark}
\newtheorem{remark}[theorem]{Remark}
\numberwithin{equation}{section}
\newcommand{\F}{\mathbb{F}}

\begin{document}
\title[New extremal binary self-dual codes from ring extensions]{Extension
Theorems for Self-dual codes over rings and new binary self-dual codes}
\author{Abidin Kaya}
\author{Bahattin Yildiz}
\address{Department of Mathematics, Fatih University, 34500, Istanbul, Turkey%
}
\email{ akaya@fatih.edu.tr, byildiz@fatih@edu.tr}

\begin{abstract}
In this work, extension theorems are generalized to self-dual codes over
rings and as applications many new binary self-dual extremal codes are found
from self-dual codes over $\mathbb{F}_{2^m}+u\mathbb{F}_{2^m}$ for $m=1,2$.
The duality and distance preserving Gray maps from $\mathbb{F}_{4}+u\mathbb{F%
}_{4}$ to $(\mathbb{F}_{2}+u\mathbb{F}_{2})^2$ and $\mathbb{F}_2^4$ are used
to obtain self-dual codes whose binary Gray images are $[64,32,12]$-extremal
self-dual. An $\mathbb{F}_2+u\mathbb{F}_2$-extension is used and as binary images, $178$
extremal binary self-dual codes of length $68$ with new weight enumerators are obtained. Especially the first examples of codes with $\gamma =3$ and
many codes with the rare $\gamma =4,6$ parameters are obtained. In addition
to these, two hundred fifty doubly even self dual $[96,48,16]$-codes with
new weight enumerators are obtained from four-circulant codes over $\mathbb{F%
}_4+u\mathbb{F}_4$. New extremal doubly even binary codes of lengths $80$
and $88$ are also found by the $\mathbb{F}_2+u\mathbb{F}_2$-lifts of binary
four circulant codes and a corresponding result about 3-designs is stated.
\end{abstract}

\maketitle

\section{Introduction}

The construction of extremal binary self-dual codes has generated a
considerable interest among researchers recently. The connection of these
codes to designs, lattices and other such mathematical objects has been a
source of motivation for this interest. Several construction methods have
been employed for this purpose. Among the most common ones, we can mention
double and bordered double-circulant constructions, constructions with a
specific automorphism group, and recently ring constructions using different
rings of characteristic $2$. We refer the reader to \cite%
{dontcheva,dontcheva2,feit,goodwin,gulliver,karadeniz,karadeniz2,Kim,yorgova}
and \cite{tsai} for more on these constructions.

Ling and Sole studied Type II codes over the ring $\mathbb{F}_{4}+u\mathbb{F}%
_{4}$ in \cite{ling}, which was later generalized to the ring $\mathbb{F}%
_{2^{m}}+u\mathbb{F}_{2^{m}}$ in \cite{betsumiya}. These rings behave
similar to the oft-studied ring $\mathbb{F}_{2}+u\mathbb{F}_{2}$ in the
literature. The common theme in the aforementioned works is that a distance
and duality preserving Gray map can be defined that takes codes over those
rings to binary codes, preserving the linearity, the weight distribution and
the duality.

Harada and Kim give two different extension methods in \cite{Harada} and
\cite{Kim} respectively for binary self-dual codes. Both methods describe
how a binary self-dual code of length $n$ can be extended to obtain a binary
self-dual code of length $n+2$.

In this work we generalize the extension methods described on the binary
field to any binary ring(i.e., a ring of characteristic 2). With this method
we extend self-dual codes over binary rings to further lengths which
correspond to a more diverse set of lengths. Also with the rich algebraic
structure of the ring, we have a better chance to get good self-dual codes.
The binary rings that we use are mainly $\mathbb{F}_{4}+u\mathbb{F}_{4}$ and
$\mathbb{F}_{2}+u\mathbb{F}_{2}$ as we already have distance and
duality-preserving Gray maps for these rings. Using these methods we were
able to obtain 178 new extremal binary self-dual codes of length 68 and 14 new extremal codes of length 80.

The rest of the paper is organized as follows: Preliminaries about codes
over $\mathbb{F}_4+u\mathbb{F}_4$ and the distance and duality-preserving
Gray maps are given in section 2. In section 3, we give constructions for
binary self-dual codes of length $64$ coming from the Gray images of
four-circulant self-dual codes over $\mathbb{F}_4+u\mathbb{F}_4$. In section
4, we describe the ring extension methods to extend self-dual codes over
binary rings. In section 5, we apply the ring extension to codes obtained in
section 3 to obtain a number of extremal binary self-dual codes of length $%
68 $ with new parameters in their weight enumerators. In section 6, we
describe constructions of extremal binary self-dual codes of length $80$ and
$88$ as well as new Type II codes of length $96$ from codes over $\mathbb{F}%
_{2^m}+u\mathbb{F}_{2^m}$ for $m=1,2$.

\section{Preliminaries}

Let $\mathbb{F}_{4}=\mathbb{F}_{2}\left( \omega \right) $ be the quadratic field extension of $\F_2$, where $\omega
^{2}+\omega +1=0$. The ring $\mathbb{F}_{4}+u\mathbb{F}_{4}$ defined via $%
u^{2}=0$ is a commutative binary ring of size $16$. We may easily observe
that it is isomorphic to $\mathbb{F}_{2}\left[ \omega ,u\right]
/\left\langle u^{2},\omega ^{2}+\omega +1\right\rangle $. The ring has a
unique non-trivial ideal $\left\langle u\right\rangle =\left\{ 0,u,u\omega
,u+u\omega \right\} $. Note that $\mathbb{F}_4+u\mathbb{F}_4$ can be viewed
as an extension of $\mathbb{F}_2+u\mathbb{F}_2$ and so we can describe any
element of $\mathbb{F}_4+u\mathbb{F}_4$ in the form $\omega a+\bar{\omega}b$
uniquely, where $a,b \in \mathbb{F}_2+u\mathbb{F}_2$.

A code $C$ of length $n$ over $\mathbb{F}_{4}+u\mathbb{F}_{4}$ is an $\left(
\mathbb{F}_{4}+u\mathbb{F}_{4}\right) $-submodule of $\left( \mathbb{F}_{4}+u%
\mathbb{F}_{4}\right) ^{n}$. Elements of the code $C$ are called codewords
of $C$. Let $x=\left( x_{1},x_{2},\ldots ,x_{n}\right) $ and $y=\left(
y_{1},y_{2},\ldots ,y_{n}\right) $ be two elements of $\left( \mathbb{F}%
_{4}+u\mathbb{F}_{4}\right) ^{n}$. The duality is understood in terms of the
Euclidean inner product; $\left\langle x,y\right\rangle _{E}=\sum x_{i}y_{i}$%
. The dual $C^{\bot }$ of the code $C$ is defined as
\begin{equation*}
C^{\bot }=\left\{ x\in \left( \mathbb{F}_{4}+u\mathbb{F}_{4}\right) ^{n}\mid
\left\langle x,y\right\rangle _{E}=0\text{ for all }y\in C\right\} .
\end{equation*}%
We say that $C$ is self-dual if $C=C^{\bot }$. Let us recall the following
Gray Maps from \cite{gaborit} and \cite{dougherty2};%
\begin{equation*}
\begin{tabular}{l||l}
$\psi _{\mathbb{F}_{4}}:\left( \mathbb{F}_{4}\right) ^{n}\rightarrow \left(
\mathbb{F}_{2}\right) ^{2n}$ & $\varphi _{\mathbb{F}_{2}+u\mathbb{F}%
_{2}}:\left( \mathbb{F}_{2}+u\mathbb{F}_{2}\right) ^{n}\rightarrow \mathbb{F}%
_{2}^{2n}$ \\
$a\omega +b\overline{\omega }\mapsto \left( a,b\right) \text{, \ }a,b\in
\mathbb{F}_{2}^{n}$ & $a+bu\mapsto \left( b,a+b\right) \text{, \ }a,b\in
\mathbb{F}_{2}^{n}.$%
\end{tabular}%
\end{equation*}%
In \cite{ling}, those were generalized to the following Gray maps;
\begin{equation*}
\begin{tabular}{l||l}
$\psi _{\mathbb{F}_{4}+u\mathbb{F}_{4}}:\left( \mathbb{F}_{4}+u\mathbb{F}%
_{4}\right) ^{n}\rightarrow \left( \mathbb{F}_{2}+u\mathbb{F}_{2}\right)
^{2n}$ & $\varphi _{\mathbb{F}_{4}+u\mathbb{F}_{4}}:\left( \mathbb{F}_{4}+u%
\mathbb{F}_{4}\right) ^{n}\rightarrow \mathbb{F}_{4}^{2n}$ \\
$a\omega +b\overline{\omega }\mapsto \left( a,b\right) \text{, \ }a,b\in
\left( \mathbb{F}_{2}+u\mathbb{F}_{2}\right) ^{n}$ & $a+bu\mapsto \left(
b,a+b\right) \text{, \ }a,b\in \mathbb{F}_{4}^{n}$%
\end{tabular}%
\end{equation*}%
Note that these Gray maps preserve orthogonality in the respective
alphabets, for the details we refer to \cite{ling}. The binary codes $%
\varphi _{\mathbb{F}_{2}+u\mathbb{F}_{2}}\circ \psi _{\mathbb{F}_{4}+u%
\mathbb{F}_{4}}\left( C\right) $ and $\psi _{\mathbb{F}_{4}}\circ \varphi _{%
\mathbb{F}_{4}+u\mathbb{F}_{4}}\left( C\right) $ are equivalent to each
other. The Lee weight of an element in $\mathbb{F}_{4}+u\mathbb{F}_{4}$ is
defined to be the Hamming weight of its binary image under any of the previously mentioned compositions of the maps. A self-dual code is
said to be of Type II if the Lee weights of all codewords are multiples of $%
4 $, otherwise it is said to be of Type I.

\begin{proposition}
$($\cite{ling}$)$ Let $C$ be a code over $\mathbb{F}_{4}+u\mathbb{F}_{4}$.
If $C$ is self-orthogonal, so are $\psi _{\mathbb{F}_{4}+u\mathbb{F}%
_{4}}\left( C\right) $ and $\varphi _{\mathbb{F}_{4}+u\mathbb{F}_{4}}\left(
C\right) $. $C$ is a Type I (resp. Type II) code over $\mathbb{F}_{4}+u%
\mathbb{F}_{4}$ if and only if $\varphi _{\mathbb{F}_{4}+u\mathbb{F}%
_{4}}\left( C\right) $ is a Type I (resp. Type II) $\mathbb{F}_{4}$-code, if
and only if $\psi _{\mathbb{F}_{4}+u\mathbb{F}_{4}}\left( C\right) $ is a
Type I (resp. Type II) $\mathbb{F}_{2}+u\mathbb{F}_{2}$-code. Furthermore,
the minimum Lee weight of $C$ is the same as the minimum Lee weight of $\psi
_{\mathbb{F}_{4}+u\mathbb{F}_{4}}\left( C\right) $ and $\varphi _{\mathbb{F}%
_{4}+u\mathbb{F}_{4}}\left( C\right) $.
\end{proposition}

\begin{corollary}
Suppose that $C$ is a self-dual code over $\mathbb{F}_{4}+u\mathbb{F}_{4}$
of length $n$ and minimum Lee distance $d$. Then $\varphi _{\mathbb{F}_{2}+u%
\mathbb{F}_{2}}\circ \psi _{\mathbb{F}_{4}+u\mathbb{F}_{4}}\left( C\right) $
is a binary $\left[ 4n,2n,d\right] $ self-dual code. Moreover, $C$ and $%
\varphi _{\mathbb{F}_{2}+u\mathbb{F}_{2}}\circ \psi _{\mathbb{F}_{4}+u%
\mathbb{F}_{4}}\left( C\right) $ have the same weight enumerator. If $C$ is
Type I (Type II), then so is $\varphi _{\mathbb{F}_{2}+u\mathbb{F}_{2}}\circ
\psi _{\mathbb{F}_{4}+u\mathbb{F}_{4}}\left( C\right) $.
\end{corollary}

An upper bound on the minimum Hamming distance of a binary self-dual code is
as follows:

\begin{theorem}
$($\cite{Rains}$)$ Let $d_{I}(n)$ and $d_{II}(n)$ be the minimum distance of
a Type I and Type II binary code of length $n$, respectively. Then
\begin{equation*}
d_{II}(n)\leq 4\lfloor \frac{n}{24}\rfloor +4
\end{equation*}%
and
\begin{equation*}
d_{I}(n)\leq \left\{
\begin{array}{ll}
4\lfloor \frac{n}{24}\rfloor +4 & \text{if $n\not\equiv 22\pmod{24}$} \\
4\lfloor \frac{n}{24}\rfloor +6 & \text{if $n\equiv 22\pmod{24}$.}%
\end{array}%
\right.
\end{equation*}
\end{theorem}

Self-dual codes meeting these bounds are called \textit{extremal}.
Throughout the text we obtain extremal Type I binary codes of lengths 64 and
68 and extremal Type II codes of lengths 80 and 88. The existence of
extremal Type II codes of length $96$ is as yet unknown. But we get Type II
codes of parameters $[96,48,16]$, which is the best known parameter at the
moment.

\section{$\left[ 64,32,12\right] _{2}$ singly-even codes as images of $%
\mathbb{F}_{4}+u\mathbb{F}_{4}$-lifts of codes over $\mathbb{F}_{4}$}

The double circulant and bordered double circulant constructions are quite
commonly used constructions in the literature for self-dual codes. However
there is a variation of these constructions called the four circulant
construction which has recently been introduced and used in the context of
self-dual codes. We will apply the construction here. The four circulant
construction was applied to the ring $\mathbb{F}_2+u\mathbb{F}_2$ in \cite%
{karadeniz2} to obtain extremal binary self-dual codes. The main theorem
that can exactly be extended to include the ring $\mathbb{F}_4+u\mathbb{F}_4$
is the following:

\begin{theorem}
\label{fourcirc}$($ \cite{karadeniz2}, with $\mathbb{F}_2$ replaced by $%
\mathbb{F}_4$ $)$ Let $C$ be the linear code over $\mathbb{F}_4+u\mathbb{F}%
_4 $ of length $4n$ generated by the four circulant matrix
\begin{equation*}
G:=\left [ \enspace I_{2n}\enspace
\begin{array}{|cc}
A & B \\
B^{T} & A^{T}%
\end{array}
\right ]
\end{equation*}
where $A$ and $B$ are circulant $n\times n$ matrices over $\mathbb{F}_4+u%
\mathbb{F}_4$ satisfying $AA^{T}+BB^{T} = I_{n}$. Then $C$ is self-dual.
\end{theorem}

The proof being exactly the same as the case of $\mathbb{F}_2+u\mathbb{F}_2$%
, is omitted here.

Now, our aim is to find extremal binary self-dual codes using the four
circulant construction over $\mathbb{F}_4+u\mathbb{F}_4$. This requires a
restriction on the minimum weight. To reduce the search field we will
consider the projection $\mu: \mathbb{F}_4+u\mathbb{F}_4 \rightarrow \mathbb{%
F}_4$ by letting $\mu(a+bu) = a$ for all $a, b \in \mathbb{F}_4$. This map
then can be extended in a natural way to $(\mathbb{F}_4+u\mathbb{F}_4)^n$.
It can easily be shown that $\mu$ preserves duality and because of the type
of the matrix, we can say that if $C$ is a four circulant self-dual code
generated by a matrix $G$ of the form given in Theorem \ref{fourcirc}, then $%
\mu(C)$ will also be a four circulant self dual code over $\mathbb{F}_4$
generated by the matrix $\mu(G)$. Thus any four circulant self-dual code
over $\mathbb{F}_4+u\mathbb{F}_4$ can be viewed as a lift of a four
circulant self-dual code over $\mathbb{F}_4$ of the same length. The
following theorem, an analogue of which can also be found in \cite%
{karadeniz2} reduces the search field quite considerably:

\begin{theorem}
\label{bound}$($ \cite{karadeniz2}, with $\mathbb{F}_2$ replaced by $\mathbb{%
F}_4$ $)$ Suppose $C$ is a linear code over $\mathbb{F}_4+u\mathbb{F}_4$ and
that $C^{\prime }= \mu(C)$ is its projection to $\mathbb{F}_4$. With $d$ and
$d^{\prime }$ representing the minimum Lee distances of $C$ and $C^{\prime }$
respectively, we have $d \leq 2d^{\prime }.$
\end{theorem}

So, to construct binary extremal self-dual codes of length $64$, we need
self-dual codes over $\mathbb{F}_{4}+u\mathbb{F}_{4}$ of length $16$ and
minimum Lee weight $12$. However the projections of four circulant self-dual
codes over $\mathbb{F}_{4}+u\mathbb{F}_{4}$ being four circulant self-dual
codes over $\mathbb{F}_{4}$, by Theorem \ref{bound} we need four circulant
self-dual codes over $\mathbb{F}_{4}$ of minimum Lee weight at least $6$. A
complete classification of all four-circulant self-dual codes over $\mathbb{F%
}_{4}$ of length $16$ can be done by considering all possible first rows for
the matrices $A$ and $B$, denoted henceforth by $r_A$ and $r_B$, which requires a search over $4^{8}$ possible
matrices, only a portion of which will be self-dual with minimum Lee weight $%
\geq 6$. Lifting these to $\mathbb{F}_{4}+u\mathbb{F}_{4}$, we see that only
the codes listed in Table 1 have resulted in self-dual codes with extremal
binary images.
\begin{table}[tbp]
\caption{Four circulant codes over $\mathbb{F}_{4}$}
\label{tab:fourcirc}
\begin{center}
\begin{tabular}{|l|l|l|l|l|}
\hline
$\mathcal{C}_{i}$ & $r_{A}$ & $r_{B}$ & $\psi_{\mathbb{F}_{4}}(C)$ & $%
\left\vert Aut\left( C\right) \right\vert $ \\ \hline
$\mathcal{C}_{1}$ & $\left( 1,\omega ,\omega ,0\right) $ & $\left( \omega
,1+\omega ,1+\omega ,\omega \right) $ & $[32,16,8]_{2}$ & $2^{12}3\times 7$
\\ \hline
$\mathcal{C}_{2}$ & $\left( 1,0,1,\omega \right) $ & $\left( 0,0,1+\omega
,0\right) $ & $[32,16,6]_{2}$ & $2^{9}3^{2}5$ \\ \hline
$\mathcal{C}_{3}$ & $\left( \omega ,\omega ,1+\omega ,1+\omega \right) $ & $%
\left( 1+\omega ,\omega ,0,0\right) $ & $[32,16,6]_{2}$ & $2^{9}3^{2}5$ \\
\hline
$\mathcal{C}_{4}$ & $\left( 1,0,1,1+\omega \right) $ & $\left( 0,0,\omega
,0\right) $ & $[32,16,6]_{2}$ & $2^{9}3^{2}5$ \\ \hline
$\mathcal{C}_{5}$ & $\left( 1+\omega ,0,1+\omega ,\omega \right) $ & $\left(
0,0,1+\omega ,0\right) $ & $[32,16,6]_{2}$ & $2^{9}3^{2}5$ \\ \hline
\end{tabular}%
\end{center}
\end{table}
There are two possibilities for the weight enumerators of extremal
singly-even $\left[ 64,32,12\right] _{2}$ codes (\cite{conway}):
\begin{eqnarray*}
W_{64,1} &=&1+\left( 1312+16\beta \right) y^{12}+\left( 22016-64\beta
\right) y^{14}+\cdots , \: 14 \leq \beta \leq 284, \\
W_{64,2} &=&1+\left( 1312+16\beta \right) y^{12}+\left( 23040-64\beta
\right) y^{14}+\cdots, \: 0\leq \beta \leq 277.
\end{eqnarray*}%
The theoretical values for $\beta$ have not all been constructed yet. Most
recently, codes with $\beta =$25, 39, 53 and 60 in $W_{64,1}$ and $\beta =$%
51 and 58 in $W_{64,2}$ are constructed in \cite{yankov}, a code with $\beta
=80$ in $W_{64,2}$ is constructed in \cite{karadeniz2}. Together with these,
codes exist with weight enumerators $\beta =$14, 18, 22, 25, 32, 36, 39, 44,
46, 53, 60 and 64 in $W_{64,1}$ and for $\beta =$0, 1, 2, 4, 5,\ 6, 8, 9,
10, 12, 13,\ 14, 16,\ 17, 18, 20, 21,\ 22, 23, 24,\ 25,\ 28,\ 19,\ 30, 32,\
33,\ 36, 37, 38, 40,\ 41,\ 44, 48, 51,\ 52,\ 56, 58, 64, 72, 80,\ 88,\ 96,
104, 108,\ 112,\ 114,\ 118,\ 120 and 184 in $W_{64,2}$.

In order to fit the upcoming tables regarding the results, we label the
elements of $\mathbb{F}_{4}+u\mathbb{F}_{4}$ as follows;

\begin{center}
\begin{tabular}{|l|l||l|l||l|l||l|l|}
\hline
$z_{1}$ & $0$ & $a_{1}$ & $1$ & $b_{1}$ & $\omega $ & $c_{1}$ & $1+\omega $
\\ \hline
$z_{2}$ & $u$ & $a_{2}$ & $1+u$ & $b_{2}$ & $\omega +u$ & $c_{2}$ & $%
1+\omega +u$ \\ \hline
$z_{3}$ & $u\omega $ & $a_{3}$ & $1+u\omega $ & $b_{3}$ & $\omega +u\omega $
& $c_{3}$ & $1+\omega +u\omega $ \\ \hline
$z_{4}$ & $u+u\omega $ & $a_{4}$ & $1+u+u\omega $ & $b_{4}$ & $\omega
+u+u\omega $ & $c_{4}$ & $1+\omega +u+u\omega $ \\ \hline
\end{tabular}
\end{center}

We lift the $\mathbb{F}_4$-codes given in Table 1 to $\mathbb{F}_{4}+u%
\mathbb{F}_4$, as a result of which we obtain extremal binary self-dual
codes of length $64$ as given in Table 2.

\begin{table}[tbp]
\caption{The $\mathbb{F}_{4}+u\mathbb{F}_{4}$-lifts of $\mathcal{C}_i$ and
the $\protect\beta$ values of the binary images}
\label{tab:64codes}
\begin{center}
\begin{tabular}{|l|l|l|l|l|l|}
\hline
code &  & first row of $A$ & first row of $B$ & $\beta $ in $W_{64,2}$ & $%
\left\vert Aut\left( C\right) \right\vert $ \\ \hline
$\mathcal{J}_{1}$ & $\mathcal{C}_{1}$ & $\left(
a_{2},b_{3},b_{1},z_{4}\right) $ & $\left( b_{4},c_{4},c_{1},b_{2}\right) $
& $48$ & $2^{5}$ \\ \hline
$\mathcal{J}_{2}$ & $\mathcal{C}_{1}$ & $\left(
a_{3},b_{2},b_{4},z_{1}\right) $ & $\left( b_{3},c_{3},c_{2},b_{1}\right) $
& $52$ & $2^{5}$ \\ \hline
$\mathcal{K}_{1}$ & $\mathcal{C}_{2}$ & $\left(
a_{4},z_{3},a_{3},b_{1}\right) $ & $\left( z_{2},z_{4},c_{3},z_{1}\right) $
& $8$ & $2^{5}$ \\ \hline
$\mathcal{K}_{2}$ & $\mathcal{C}_{2}$ & $\left(
a_{1},z_{4},a_{2},b_{4}\right) $ & $\left( z_{2},z_{4},c_{4},z_{1}\right) $
& $40$ & $2^{5}$ \\ \hline
$\mathcal{K}_{3}$ & $\mathcal{C}_{2}$ & $\left(
a_{2},z_{4},a_{1},b_{4}\right) $ & $\left( z_{2},z_{4},c_{1},z_{1}\right) $
& $40$ & $2^{5}$ \\ \hline
$\mathcal{K}_{4}$ & $\mathcal{C}_{2}$ & $\left(
a_{2},z_{2},a_{1},b_{2}\right) $ & $\left( z_{4},z_{3},c_{1},z_{2}\right) $
& $40$ & $2^{5}$ \\ \hline
$\mathcal{K}_{5}$ & $\mathcal{C}_{2}$ & $\left(
a_{2},z_{4},a_{1},b_{2}\right) $ & $\left( z_{2},z_{1},c_{1},z_{4}\right) $
& $40$ & $2^{5}$ \\ \hline
$\mathcal{L}_{1}$ & $\mathcal{C}_{3}$ & $\left(
b_{2},b_{4},c_{2},c_{3}\right) $ & $\left( c_{1},b_{1},z_{2},z_{2}\right) $
& $0$ & $2^{6}$ \\ \hline
$\mathcal{L}_{2}$ & $\mathcal{C}_{3}$ & $\left(
b_{1},b_{4},c_{1},c_{3}\right) $ & $\left( c_{2},b_{2},z_{1},z_{1}\right) $
& $0$ & $2^{5}$ \\ \hline
$\mathcal{L}_{3}$ & $\mathcal{C}_{3}$ & $\left(
b_{1},b_{3},c_{1},c_{4}\right) $ & $\left( c_{1},b_{2},z_{2},z_{1}\right) $
& $4$ & $2^{5}$ \\ \hline
$\mathcal{L}_{4}$ & $\mathcal{C}_{3}$ & $\left(
b_{4},b_{2},c_{2},c_{1}\right) $ & $\left( c_{2},b_{1},z_{1},z_{3}\right) $
& $8$ & $2^{4}$ \\ \hline
$\mathcal{L}_{5}$ & $\mathcal{C}_{3}$ & $\left(
b_{1},b_{4},c_{2},c_{3}\right) $ & $\left( c_{1},b_{3},z_{3},z_{4}\right) $
& $12$ & $2^{4}$ \\ \hline
$\mathcal{L}_{6}$ & $\mathcal{C}_{3}$ & $\left(
b_{2},b_{4},c_{3},c_{2}\right) $ & $\left( c_{3},b_{1},z_{1},z_{4}\right) $
& $24$ & $2^{5}$ \\ \hline
$\mathcal{L}_{7}$ & $\mathcal{C}_{3}$ & $\left(
b_{1},b_{4},c_{1},c_{2}\right) $ & $\left( c_{3},b_{1},z_{2},z_{2}\right) $
& $28$ & $2^{4}$ \\ \hline
$\mathcal{L}_{8}$ & $\mathcal{C}_{3}$ & $\left(
b_{2},b_{2},c_{1},c_{1}\right) $ & $\left( c_{3},b_{1},z_{1},z_{2}\right) $
& $36$ & $2^{5}$ \\ \hline
$\mathcal{M}_{1}$ & $\mathcal{C}_{4}$ & $\left(
a_{1},z_{4},a_{2},c_{1}\right) $ & $\left( z_{1},z_{4},b_{2},z_{2}\right) $
& $40$ & $2^{5}$ \\ \hline
$\mathcal{M}_{2}$ & $\mathcal{C}_{4}$ & $\left(
a_{1},z_{4},a_{2},c_{1}\right) $ & $\left( z_{1},z_{4},b_{4},z_{2}\right) $
& $40$ & $2^{5}$ \\ \hline
$\mathcal{M}_{3}$ & $\mathcal{C}_{4}$ & $\left(
a_{1},z_{4},a_{2},c_{1}\right) $ & $\left( z_{1},z_{4},b_{2},z_{2}\right) $
& $40$ & $2^{5}$ \\ \hline
$\mathcal{N}_{1}$ & $\mathcal{C}_{5}$ & $\left(
c_{1},z_{1},c_{4},b_{1}\right) $ & $\left( z_{1},z_{2},c_{2},z_{4}\right) $
& $32$ & $2^{5}$ \\ \hline
\end{tabular}%
\end{center}
\end{table}

\section{Extension methods for self-dual codes over binary rings}

In the sequel, let $S$ be a commutative ring of characteristic $2$ with
identity.

\begin{theorem}
\label{extensionA} Let $C$ be a self-dual code over $S$ of length $n$ and $%
G=(r_{i})$ be a $k\times n$ generator matrix for $C$, where $r_{i}$ is the $%
i $-th row of $G$, $1\leq i\leq k$. Let $c$ be a unit in $S$ such that $%
c^{2}=1 $ and $X$ be a vector in $S^{n}$ with $\left\langle X,X\right\rangle
=1$. Let $y_{i}=\left\langle r_{i},X\right\rangle $ for $1\leq i\leq k$.
Then the following matrix%
\begin{equation*}
\left[
\begin{array}{cc|c}
1 & 0 & X \\ \hline
y_{1} & cy_{1} & r_{1} \\
\vdots & \vdots & \vdots \\
y_{k} & cy_{k} & r_{k}%
\end{array}%
\right] ,
\end{equation*}%
generates a self-dual code $D$ over $S$ of length $n+2$.
\end{theorem}

A more specific extension method which can easily be applied to circulant
codes may be given as follows:

\begin{theorem}
\label{extensionid} Let $C$ be a self-dual code generated by $G=\left[
I_{n}|A\right] $ over $S$. If the sum of the elements in $i$-th row of $A$
is $r_{i}$ then the matrix:%
\begin{equation*}
G^{\ast }=\left[
\begin{array}{cc|cccccc}
1 & 0 & x_{1} & \ldots & x_{n} & 1 & \ldots & 1 \\ \hline
y_{1} & cy_{1} & \multicolumn{3}{c}{} & \multicolumn{3}{c}{} \\
\vdots & \vdots &  & I_{n} &  &  & A &  \\
y_{n} & cy_{n} &  &  &  &  &  &
\end{array}%
\right] ,
\end{equation*}%
where $y_{i}=x_{i}+r_{i}$, $c$ is a unit with $c^{2}=1$, $\left\langle
X,X\right\rangle =1+n$ and $X=\left( x_{1},\ldots ,x_{n}\right) $, generates
a self-dual code $C^{\ast }$ over $S$.
\end{theorem}

\begin{remark}
As can be seen, these extension theorems generalize the binary extension theorems given in \cite{Harada} and \cite{Kim}. The proofs being exactly analogous, have been omitted here.
\end{remark}
\section{New extremal binary self dual codes of length $68$ from $\mathbb{F}%
_{2}+u\mathbb{F}_{2}$ extensions}

The weight enumerator of a self-dual $\left[ 68,34,12\right] _{2}$ code is
in one of the following forms (\cite{dougherty1}):
\begin{eqnarray*}
W_{68,1} &=&1+\left( 442+4\beta \right) y^{12}+\left( 10864-8\beta \right)
y^{14}+\cdots , \\
W_{68,2} &=&1+\left( 442+4\beta \right) y^{12}+\left( 14960-8\beta
-256\gamma \right) y^{14}+\cdots
\end{eqnarray*}%
where $\beta $ and $\gamma $ are parameters. Tsai et al. constructed a
substantial number of codes in both possible weight enumerators in \cite%
{tsai}. Recently, 32 new codes are obtained in \cite{kaya} and $28$ new
codes including the first examples with $\gamma =4$ and $\gamma =6$ in $%
W_{68,2}$ are obtained in \cite{karadeniz}. Together with the ones in \cite%
{karadeniz, kaya} codes exists for $W_{68,2}$ when $\gamma =0$ and $\beta =$%
38, 40, 44, 45, 47,...,136, 138, 139, 170, 204, 238, 272; $\gamma =1$ and $%
\beta =$61, 63, 64, 65, 72, 73, 76, 77, 79, 81,\ldots , 115, 118, 126, 129,
132, 133, 138, 140, 142, 146; $\gamma =2$ and $\beta =$65, 71, 77, 82, 84,
86, 88, 93, 94, 96, 99, 109, 123, 130, 132, 134, 140, 142, 146, 152 or $%
\beta \in \left\{ 2m|51\leq m\leq 63\right\} $; $\gamma =4$ and $\beta =$%
116, 122, 124, 128, 140, 142, 152 and $\gamma =6$ with $\beta =$176. The extension
methods in section 4 are applied to the $\psi _{\mathbb{F}_{4}+u\mathbb{F}%
_{4}}$-images of the codes in table \ref{tab:64codes}. Throughout the tables %
\ref{tab:gamma0}-\ref{tab:gamma4}, \ref{tab:extA} the codes are generated over $\mathbb{F}%
_{2}+u\mathbb{F}_{2}$ by the matrices of the following form;
\begin{equation*}
\left[
\begin{array}{cc|c}
1 & 0 & X \\ \hline
y_{1} & cy_{1} &  \\
\vdots  & \vdots  & \psi _{\mathbb{F}_{4}+u\mathbb{F}_{4}}\left(
C_{i}\right)  \\
y_{k} & cy_{k} &
\end{array}%
\right] .
\end{equation*}%
The second extension theorem is used to obtain the results tabulated in Table \ref{tab:extB}.
As binary images of all these codes we were able to obtain 181 new codes in $%
W_{68,2}$, which are listed in the tables \ref{tab:gamma0}-\ref{tab:extA}.
More precisely, 14 codes with $\gamma =0$ in tables \ref{tab:gamma0},\ref%
{tab:extB} and \ref{tab:extA}, 47 codes with $\gamma =1$ listed in tables %
\ref{tab:gamma1},\ref{tab:extB} and \ref{tab:extA}, 42 codes with $\gamma =2$
in Table \ref{tab:gamma2}, 37 codes with $\gamma =3$ in tables \ref%
{tab:gamma3},\ref{tab:extB} and \ref{tab:extA} 21 codes with $\gamma =4$
listed in Table \ref{tab:gamma4} and 5 codes with $\gamma =6$ which are
listed in Table \ref{tab:gamma0}. In order to save space $1+u$ in $X$ are
replaced by 3 in tables. Note that the codes with $\gamma=3$ in their weight enumerators are the first examples in the literature of that parameter.

\begin{table}[tbp]
\caption{$\left[ 68,34,12\right] $ codes with $\protect\gamma =0$ and $%
\protect\gamma =6$ in $W_{68,2}$ (16 codes)}
\label{tab:gamma0}
\begin{center}
\begin{tabular}{|l|l|l|l|l|}
\hline
Code & $X$ & $c$ & $\gamma $ & $\beta $ \\ \hline
$\mathcal{L}_{3}$ & $[111013303300031u31u3uu10uuu000u1]$ & $1+u$ & $0$ & $46$
\\ \hline
$\mathcal{M}_{1}$ & $[u13u3u33110uu11u10110000uu13u100]$ & $1$ & $0$ & $137$
\\ \hline
$\mathcal{M}_{1}$ & $[3uu031uu30u13u3u0u31u33111113013]$ & $1+u$ & $0$ & $%
141 $ \\ \hline
$\mathcal{K}_{3}$ & $[u1030u3u1u1u033301uu333101u30101]$ & $1$ & $0$ & $142$
\\ \hline
$\mathcal{J}_{2}$ & $[0u1uuuu11u331013u1u130u113u131uu]$ & $1$ & $0$ & $143$
\\ \hline
$\mathcal{M}_{1}$ & $[1300uu030u03310031130u30000u303u]$ & $1$ & $0$ & $145$
\\ \hline
$\mathcal{J}_{2}$ & $[u11u000u31uu00030u33100303100103]$ & $1+u$ & $0$ & $%
147 $ \\ \hline
$\mathcal{N}_{1}$ & $[303010u310u010u0u01031131001033u]$ & $1$ & $0$ & $148$
\\ \hline
$\mathcal{J}_{2}$ & $[1001uu1103111u3031013113001130u0]$ & $1$ & $0$ & $149$
\\ \hline
$\mathcal{J}_{2}$ & $[3u1u033u10310330011u031003010130]$ & $1$ & $0$ & $151$
\\ \hline
$\mathcal{J}_{1}$ & $[03013310u03u313330011u0u13113030]$ & $1$ & $0$ & $153$
\\ \hline
$\mathcal{L}_{6}$ & $[101110u1301303u033311033u033uu30]$ & $1+u$ & $6$ & $%
138 $ \\ \hline
$\mathcal{L}_{6}$ & $[uu1u1u11301313uu03331101303u3uu1]$ & $1$ & $6$ & $154$
\\ \hline
$\mathcal{L}_{6}$ & $[30111uuuu0033010330003301301010u]$ & $1$ & $6$ & $156$
\\ \hline
$\mathcal{L}_{6}$ & $[30u30133113uuuu3u3u0u111u3300101]$ & $1$ & $6$ & $158$
\\ \hline
$\mathcal{L}_{6}$ & $[uu311333001u033u010110011000u131]$ & $1$ & $6$ & $162$
\\ \hline
\end{tabular}%
\end{center}
\end{table}

\begin{table}[tbp]
\caption{$\left[ 68,34,12\right] $ codes with $\protect\gamma =1$ in $%
W_{68,2}$ (30 codes)}
\label{tab:gamma1}
\begin{center}
\begin{tabular}{|l|l|l|l|}
\hline
Code & $X$ & $c$ & $\beta $ \\ \hline
$\mathcal{L}_{2}$ & $[13u30uuu3u10uuu33111311uuu010uu1]$ & $1$ & $54$ \\
\hline
$\mathcal{L}_{2}$ & $[011u031u113311uu13u310130u033u01]$ & $1$ & $56$ \\
\hline
$\mathcal{L}_{1}$ & $[1130u311u33uu31u3uu01103311uu031]$ & $1$ & $58$ \\
\hline
$\mathcal{L}_{1}$ & $[310133uuu1uu310u330uuu0101000u1u]$ & $1+u$ & $60$ \\
\hline
$\mathcal{L}_{2}$ & $[10u0033u03331131000uu033u11133uu]$ & $1+u$ & $62$ \\
\hline
$\mathcal{L}_{1}$ & $[333u111131uuu0u011uu0uu1uu1uu301]$ & $1$ & $66$ \\
\hline
$\mathcal{L}_{1}$ & $[u3u003u1u11u1uu1u101310u00003001]$ & $1+u$ & $68$ \\
\hline
$\mathcal{L}_{1}$ & $[31100103u1u10313u0u01u1u3101u033]$ & $1$ & $70$ \\
\hline
$\mathcal{L}_{1}$ & $[303010033030110uuu003uu10313uu11]$ & $1+u$ & $74$ \\
\hline
$\mathcal{L}_{3}$ & $[0u1u313u1u30033031u3311u10u30u01]$ & $1$ & $75$ \\
\hline
$\mathcal{L}_{1}$ & $[113u330u133010u3111uu1110103uu10]$ & $1$ & $78$ \\
\hline
$\mathcal{L}_{1}$ & $[030u011331000000113uu13303003131]$ & $1+u$ & $80$ \\
\hline
$\mathcal{N}_{1}$ & $[3101u3311133uu001001313u11311uu3]$ & $1$ & $119$ \\
\hline
$\mathcal{M}_{1}$ & $[0011u3uu013u3000101u01130u1101u1]$ & $1$ & $120$ \\
\hline
$\mathcal{N}_{1}$ & $[0u33103001u31u101310333011101111]$ & $1$ & $123$ \\
\hline
$\mathcal{N}_{1}$ & $[131u1u03u111u310u0u10333u3uuu033]$ & $1$ & $125$ \\
\hline
$\mathcal{N}_{1}$ & $[u303uu3313u3uu0003331011uu1u0313]$ & $1+u$ & $135$ \\
\hline
$\mathcal{N}_{1}$ & $[30u0uuu0u11330130u1311303uu1003u]$ & $1+u$ & $137$ \\
\hline
$\mathcal{K}_{2}$ & $[33u00313uu1u1uuu11uu3u00u100u11u]$ & $1+u$ & $139$ \\
\hline
$\mathcal{M}_{1}$ & $[133u030u300uu031u03u1333u0111031]$ & $1$ & $141$ \\
\hline
$\mathcal{K}_{2}$ & $[uu3u030u03uu01u1u01110u111u33000]$ & $1+u$ & $143$ \\
\hline
$\mathcal{J}_{2}$ & $[113300311310010013300u1000uu0100]$ & $1+u$ & $144$ \\
\hline
$\mathcal{K}_{2}$ & $[01103u3uu3110uu003u1u1u3300uu111]$ & $1+u$ & $145$ \\
\hline
$\mathcal{J}_{2}$ & $[uuu33uu303100101111u10uu03u33u3u]$ & $1$ & $147$ \\
\hline
$\mathcal{M}_{1}$ & $[31u013u1130031111uu00uu1001u1013]$ & $1$ & $149$ \\
\hline
$\mathcal{M}_{1}$ & $[3uu3u3u03uu0uu03330u1uu11u30u033]$ & $1+u$ & $150$ \\
\hline
$\mathcal{J}_{2}$ & $[1u03u0u33133333133u3u11u1001111u]$ & $1$ & $151$ \\
\hline
$\mathcal{K}_{2}$ & $[u1u13u0311033uu11010u1031u013u30]$ & $1+u$ & $153$ \\
\hline
$\mathcal{J}_{2}$ & $[u01u33u1133u03uu13311u31u1u3uu11]$ & $1$ & $155$ \\
\hline
$\mathcal{J}_{1}$ & $[3u3u1331u30113313133u300110u1131]$ & $1+u$ & $159$ \\
\hline
\end{tabular}%
\end{center}
\end{table}
\begin{table}[tbp]
\caption{$\left[ 68,34,12\right] $ codes with $\protect\gamma =2$ in $%
W_{68,2}$ (42 codes)}
\label{tab:gamma2}
\begin{center}
\begin{tabular}{|l|l|l|l|}
\hline
Code & $X$ & $c$ & $\beta $ \\ \hline
$\mathcal{L}_{1}$ & $[0u13u10uu131u311010001011uu30331]$ & $1+u$ & $68$ \\
\hline
$\mathcal{L}_{2}$ & $[000u3013303uuu11131u3u10uu013u30]$ & $1+u$ & $74$ \\
\hline
$\mathcal{L}_{1}$ & $[1303301100013333uu303131u11uu300]$ & $1+u$ & $76$ \\
\hline
$\mathcal{L}_{1}$ & $[u11033001u1u01u13u1u03001u1u11u0]$ & $1+u$ & $78$ \\
\hline
$\mathcal{L}_{1}$ & $[0011u003u013uu0031303uu3100u0130]$ & $1+u$ & $80$ \\
\hline
$\mathcal{L}_{3}$ & $[1uu1313u01u3u1uu111u11u31u110uuu]$ & $1$ & $85$ \\
\hline
$\mathcal{L}_{3}$ & $[30u30303u33111303u10301u3000103u]$ & $1$ & $87$ \\
\hline
$\mathcal{L}_{4}$ & $[0u1uu003010u3u311u31uuu10u0uu003]$ & $1$ & $89$ \\
\hline
$\mathcal{L}_{1}$ & $[0030113103u331000uu0133u130033u3]$ & $1$ & $90$ \\
\hline
$\mathcal{L}_{4}$ & $[0u0uuu0113303u3u1uu110u1u313u1u1]$ & $1$ & $91$ \\
\hline
$\mathcal{L}_{1}$ & $[u0113u11uuu33030313u11u1011uuu1u]$ & $1+u$ & $92$ \\
\hline
$\mathcal{L}_{3}$ & $[3110u0000113u0310133uu0u3u031u03]$ & $1+u$ & $95$ \\
\hline
$\mathcal{L}_{3}$ & $[010uu3u313300u31uuu031100u01131u]$ & $1$ & $97$ \\
\hline
$\mathcal{L}_{1}$ & $[030010113001u030u11u10u310300u31]$ & $1$ & $98$ \\
\hline
$\mathcal{L}_{1}$ & $[0011uu31333u013033u1310011011u01]$ & $1+u$ & $100$ \\
\hline
$\mathcal{L}_{7}$ & $[100u13u03uuu3303131uu033311u3313]$ & $1+u$ & $101$ \\
\hline
$\mathcal{L}_{4}$ & $[301u331013100330003u0131030330uu]$ & $1$ & $103$ \\
\hline
$\mathcal{L}_{3}$ & $[101u13u310u133u000u0u1u133331u01]$ & $1+u$ & $105$ \\
\hline
$\mathcal{N}_{1}$ & $[01u10u3333013u3u01030u0uuuu33uuu]$ & $1$ & $111$ \\
\hline
$\mathcal{N}_{1}$ & $[33u3u0u11uu3uu0u00u00u1u01u331uu]$ & $1+u$ & $115$ \\
\hline
$\mathcal{L}_{4}$ & $[3101u103u0100uu1u001133u13011130]$ & $1$ & $117$ \\
\hline
$\mathcal{N}_{1}$ & $[u130u031uu10u3101u0031131u0u1001]$ & $1$ & $119$ \\
\hline
$\mathcal{L}_{4}$ & $[11313u1111131131u0uu3u0033u03uu0]$ & $1+u$ & $121$ \\
\hline
$\mathcal{N}_{1}$ & $[011131u0u0u0300001333u33u10u3uu3]$ & $1+u$ & $125$ \\
\hline
$\mathcal{M}_{1}$ & $[u0uu11u1010u1u1u33010u3uu0u00131]$ & $1+u$ & $127$ \\
\hline
$\mathcal{M}_{1}$ & $[01u013u3311130011030u30031uu01u3]$ & $1$ & $128$ \\
\hline
$\mathcal{M}_{1}$ & $[000uuu0u331010010u301u0u0101000u]$ & $1+u$ & $129$ \\
\hline
$\mathcal{K}_{4}$ & $[0u00u00333013u30001010u0u1110011]$ & $1$ & $131$ \\
\hline
$\mathcal{M}_{1}$ & $[u03u31u31u0u03u0u103u31111uuuu13]$ & $1+u$ & $133$ \\
\hline
$\mathcal{M}_{1}$ & $[u1030133u033311113u03u0101uu3130]$ & $1+u$ & $135$ \\
\hline
$\mathcal{K}_{3}$ & $[13001uu31u1310u1u31u0031u101u031]$ & $1$ & $136$ \\
\hline
$\mathcal{K}_{3}$ & $[113uu300u0331u0u0u3101u130u1u103]$ & $1+u$ & $137$ \\
\hline
$\mathcal{K}_{4}$ & $[030u1130u10u0111uu30u1u000133011]$ & $1$ & $139$ \\
\hline
$\mathcal{N}_{1}$ & $[u01u0u0u1uuu33131u13u3u013033311]$ & $1$ & $144$ \\
\hline
$\mathcal{J}_{2}$ & $[1303111111u33u01301u30031u0000uu]$ & $1$ & $145$ \\
\hline
$\mathcal{N}_{1}$ & $[0u311301u0103u3103u3013uuuuu101u]$ & $1+u$ & $148$ \\
\hline
$\mathcal{K}_{3}$ & $[3133u1u3uu01333u0303uu3u30uu10u0]$ & $1+u$ & $150$ \\
\hline
$\mathcal{M}_{1}$ & $[03010uuu0u3131uu03uu0u0033011130]$ & $1$ & $151$ \\
\hline
$\mathcal{K}_{4}$ & $[0u331100130u111330303u3033u3101u]$ & $1$ & $153$ \\
\hline
$\mathcal{K}_{3}$ & $[0uuu03u303130303uu0301uu33uuu0]$ & $1$ & $155$ \\
\hline
$\mathcal{K}_{3}$ & $[u30u1u301330030103u0u1003u1u1103]$ & $1$ & $158$ \\
\hline
$\mathcal{L}_{8}$ & $[33uu113uu00031u30u3333u031001uu0]$ & $1$ & $160$ \\
\hline
$\mathcal{M}_{1}$ & $[3303133u1u1u30uu111003uu010u1uuu]$ & $1$ & $162$ \\
\hline
\end{tabular}%
\end{center}
\end{table}
\begin{table}[tbp]
\caption{$\left[ 68,34,12\right] $ codes with $\protect\gamma =3$ in $%
W_{68,2}$ (34 codes)}
\label{tab:gamma3}
\begin{center}
\begin{tabular}{|l|l|l|l|}
\hline
Code & $X$ & $c$ & $\beta $ \\ \hline
$\mathcal{L}_{2}$ & $[11101uu0uu113uu001u001u3u0311301]$ & $1$ & $88$ \\
\hline
$\mathcal{L}_{2}$ & $[3033311u3uuu31uu301u3u1310u00013]$ & $1+u$ & $90$ \\
\hline
$\mathcal{L}_{2}$ & $[u330u001013u33u3u333u3101303010u]$ & $1$ & $96$ \\
\hline
$\mathcal{L}_{3}$ & $[301113uu10110u3u011uuu00333uuuu1]$ & $1$ & $100$ \\
\hline
$\mathcal{L}_{2}$ & $[33u03331u3u1u010031u3333uu3111uu]$ & $1+u$ & $102$ \\
\hline
$\mathcal{L}_{2}$ & $[30u1u13u000110uuu3u3u113010u1301]$ & $1$ & $104$ \\
\hline
$\mathcal{L}_{1}$ & $\left[ 3u313u3u3u00133u010013100u011u33\right] $ & $1+u$
& $108$ \\ \hline
$\mathcal{K}_{1}$ & $[1u11030uu3303111u3uu03u3100u0030]$ & $1$ & $112$ \\
\hline
$\mathcal{L}_{3}$ & $[u1u00uu33u33330uu01uuu133013u1u1]$ & $1+u$ & $114$ \\
\hline
$\mathcal{L}_{3}$ & $[3u03u01010003u0u0u1303uu0u331000]$ & $1$ & $116$ \\
\hline
$\mathcal{L}_{4}$ & $[u03u330uu331303uu0301u0311u3333u]$ & $1+u$ & $117$ \\
\hline
$\mathcal{L}_{5}$ & $[031u0030u030u013u1u311u111303u33]$ & $1+u$ & $126$ \\
\hline
$\mathcal{M}_{1}$ & $[31103001113313101uuu1uu13031u10u]$ & $1+u$ & $127$ \\
\hline
$\mathcal{L}_{6}$ & $[1uu10uuu30133010113uu33303011113]$ & $1$ & $128$ \\
\hline
$\mathcal{L}_{5}$ & $[303u00uuu13033uu113u3313011u1uu1]$ & $1$ & $130$ \\
\hline
$\mathcal{N}_{1}$ & $[u033301u1u311313133uu31133010030]$ & $1+u$ & $133$ \\
\hline
$\mathcal{L}_{7}$ & $[31u313u0u0u131u31300u3u3u0u3uuu3]$ & $1$ & $136$ \\
\hline
$\mathcal{M}_{3}$ & $[0u100031u010uu331111u0u0u100u000]$ & $1+u$ & $137$ \\
\hline
$\mathcal{L}_{7}$ & $[33331u033u1u03u0110u1u1uu3u03u33]$ & $1$ & $138$ \\
\hline
$\mathcal{L}_{7}$ & $[3uu1333110130uuu01uu0u113310110u]$ & $1$ & $140$ \\
\hline
$\mathcal{M}_{1}$ & $[33u11100330u133001u031u00301u110]$ & $1$ & $141$ \\
\hline
$\mathcal{L}_{6}$ & $[u1u301u30u03u1u00103011310313u00]$ & $1+u$ & $142$ \\
\hline
$\mathcal{J}_{2}$ & $[u0030u11100uu30uu1u13u00uu311303]$ & $1$ & $144$ \\
\hline
$\mathcal{M}_{1}$ & $[33u11u3103333uu330031u00310uu3uu]$ & $1+u$ & $145$ \\
\hline
$\mathcal{M}_{1}$ & $[00300uuu111311uu0300100uu001uu1u]$ & $1+u$ & $147$ \\
\hline
$\mathcal{K}_{5}$ & $[03000013u0u133u030u0uu3131131300]$ & $1$ & $148$ \\
\hline
$\mathcal{M}_{2}$ & $[01331u113u0u3331000uu3u11103u3u0]$ & $1$ & $149$ \\
\hline
$\mathcal{K}_{4}$ & $[3uu300u0uu0310u031131u01010u11u0]$ & $1$ & $153$ \\
\hline
$\mathcal{K}_{5}$ & $[u3u0u133001uu13311u01u1001111111]$ & $1$ & $154$ \\
\hline
$\mathcal{K}_{2}$ & $[1uu13u3011300u3u3110u03u03311u10]$ & $1+u$ & $158$ \\
\hline
$\mathcal{M}_{2}$ & $[1uuu33001u03303033uu10u3101u00uu]$ & $1+u$ & $159$ \\
\hline
$\mathcal{K}_{3}$ & $[103u3010u11uu1u133111033u0u13310]$ & $1$ & $160$ \\
\hline
$\mathcal{J}_{1}$ & $[1001133uu3013u1010031u311u30uuu3]$ & $1$ & $162$ \\
\hline
$\mathcal{J}_{2}$ & $[u0uu1u10u00330u00u0uu0u100u33330]$ & $1+u$ & $193$ \\
\hline
\end{tabular}%
\end{center}
\end{table}
\begin{table}[tbp]
\caption{$\left[ 68,34,12\right] $ codes with $\protect\gamma =4$ in $%
W_{68,2}$ (21 codes)}
\label{tab:gamma4}
\begin{center}
\begin{tabular}{|l|l|l|l|}
\hline
Code & $X$ & $c$ & $\beta $ \\ \hline
$\mathcal{L}_{2}$ & $[11031001u00uu0u13u01u10u1u333u31]$ & $1$ & $102$ \\
\hline
$\mathcal{L}_{1}$ & $[11u11111u0u3uuuuu110111310u03010]$ & $1$ & $110$ \\
\hline
$\mathcal{L}_{2}$ & $[301u131uu3u1133311303u1uu13uu3uu]$ & $1+u$ & $120$ \\
\hline
$\mathcal{L}_{1}$ & $[1u100313uu3001311u0u01u30131uu33]$ & $1+u$ & $130$ \\
\hline
$\mathcal{L}_{5}$ & $[u110uu00u3u01113103u11u00u3030u0]$ & $1$ & $134$ \\
\hline
$\mathcal{L}_{5}$ & $[u10003uu100u03031u00013333u0u1u1]$ & $1+u$ & $136$ \\
\hline
$\mathcal{L}_{6}$ & $[u310u000uuu0uu3u101u33111u33003u]$ & $1+u$ & $138$ \\
\hline
$\mathcal{L}_{6}$ & $[u131u110u130u0u013101u00u1100uu0]$ & $1+u$ & $150$ \\
\hline
$\mathcal{K}_{2}$ & $[30uu33000u013u30303u13u303033u03]$ & $1$ & $154$ \\
\hline
$\mathcal{K}_{3}$ & $[3u33u0u30uu13u003100130311u10u30]$ & $1+u$ & $156$ \\
\hline
$\mathcal{K}_{2}$ & $[101u010u300u303u3uuuu1u11113u130]$ & $1$ & $158$ \\
\hline
$\mathcal{K}_{3}$ & $[10310uu1133u31030331u030010u01u3]$ & $1$ & $160$ \\
\hline
$\mathcal{K}_{2}$ & $[131110013uu13uu300u300001313u013]$ & $1+u$ & $162$ \\
\hline
$\mathcal{K}_{2}$ & $[0u311301u01131u103u30111003u0311]$ & $1+u$ & $164$ \\
\hline
$\mathcal{K}_{5}$ & $[3301u3133u01u33uu3013u3u31u01u13]$ & $1+u$ & $166$ \\
\hline
$\mathcal{K}_{3}$ & $[uu0u0u301u1111313u1uuu3u11u110u0]$ & $1$ & $168$ \\
\hline
$\mathcal{K}_{3}$ & $[11u0133133u0uu3313u1u0uu0u33330u]$ & $1$ & $170$ \\
\hline
$\mathcal{K}_{5}$ & $[1130uu10003313113u1uu1uu300uuuu3]$ & $1+u$ & $172$ \\
\hline
$\mathcal{K}_{3}$ & $[001u013100301u11u0313003uuu33100]$ & $1$ & $174$ \\
\hline
$\mathcal{K}_{2}$ & $[31u30033u03u033u0101u0u11111301u]$ & $1+u$ & $176$ \\
\hline
$\mathcal{K}_{3}$ & $[3301uu0u30001u1uu3u33313u1031uuu]$ & $1+u$ & $180$ \\
\hline
\end{tabular}%
\end{center}
\end{table}

\subsection{New codes from a previously constructed code}

Karadeniz et al. constructed four circulant codes of length $32$ over $%
\mathbb{F}_{2}+u\mathbb{F}_{2}$ wwhose Gray images are extremal singly-even
binary codes of length $64$ in \cite{karadeniz2}. One of these codes has a
new weight enumerator in $W_{64,2}$ with $\beta =80$. Since the $\beta $%
-value of this code is greater than that of the codes we were able to
construct, we apply the extension methods to this code. We were able to
obtain a substantial number of binary extremal codes of length $68$ with new
weight enumerators in $W_{68,2}$ as Gray images of $\mathbb{F}_{2}+u\mathbb{F%
}_{2}$-extensions.

Let $\mathcal{C}_{64}$ be the four circulant code over $\mathbb{F}_{2}+u%
\mathbb{F}_{2}$ with $r_{A}=\left( u,0,0,0,u,1,u,1+u\right) $ and $%
r_{B}=\left( u,u,0,1,1,1+u,1+u,1+u\right) $. The extension method in Theorem %
\ref{extensionid} is applied to $\mathcal{C}_{64}$ and 24 new codes in $%
W_{68,2}$ are obtained as Gray images of the extensions, so the codes in
Table \ref{tab:extB} are the Gray images of the codes generated by;
\begin{equation*}
\left[
\begin{array}{cc|cccccc}
1 & 0 & x_{1} & \ldots & x_{16} & 1 & \ldots & 1 \\ \hline
y_{1} & cy_{1} & \multicolumn{3}{c}{} & \multicolumn{3}{c}{} \\
\vdots & \vdots &  & I_{16} &  &  & M &  \\
y_{16} & cy_{16} &  &  &  &  &  &
\end{array}%
\right]
\end{equation*}%
where $M$ is the four-circulant matrix corresponding to $\mathcal{C}_{64}$
and $X=\left( x_{1},x_{2},\ldots ,x_{16}\right) $ is a random vector over $%
\mathbb{F}_{2}+u\mathbb{F}_{2}$ which satisfies $\left\langle
X,X\right\rangle =1$ and $y_{i}=x_{i}+1+u$.
\begin{table}[tbp]
\caption{$\left[ 68,34,12\right] $-codes in $W_{68,2}$ by Theorem \protect
\ref{extensionid} (22  codes)}
\label{tab:extB}
\begin{center}
\begin{tabular}{|l|l|l|l||l|l|l|l|}
\hline
$X$ & $c$ & $\gamma $ & $\beta $ & $X$ & $c$ & $\gamma $ & $\beta $ \\ \hline
 $\lbrack 3u3uu3310010u3u0]$ & $1+u$ & $0$ & $160$& $[0u013u3u0000u303]$ & $1$
& $0$ & $164$ \\ \hline
$\lbrack 1uu3313331u001uu]$ & $1+u$ & $0$ & $162$ & $[30u33u313uu300u0]$ & $%
1+u$ & $0$ & $166$ \\ \hline
$\lbrack u13310u0u1100u1u]$ & $1$ & $0$ & $140$ & $[01u33u333u3u3300]$ & $1$
& $0$ & $168$ \\ \hline
$\lbrack 031u133101uu31u0]$ & $1+u$ & $0$ & $144$ & $[1010uu30330303u0]$ & $%
1+u$ & $1$ & $154$ \\ \hline
$\lbrack 113010u1u3001130]$ & $1$ & $0$ & $146$ & $[u330uu13uu100uuu]$ & $1$
& $1$ & $156$ \\ \hline
$\lbrack 1u00311u1u131030]$ & $1+u$ & $0$ & $150$ & $[uuu01000303uuu11]$ & $%
1+u$ & $1$ & $158$ \\ \hline
$\lbrack 311333300303u13u]$ & $1$ & $0$ & $152$ & $[0u010uu0130u0310]$ & $%
1+u $ & $1$ & $160$ \\ \hline
$\lbrack 0u03000013u03300]$ & $1$ & $0$ & $154$ & $[1330u001uuuu1013]$ & $1$
& $1$ & $162$ \\ \hline
$\lbrack 3u310u1311130u11]$ & $1+u$ & $0$ & $156$ & $[uuu10110uu0uu103]$ & $%
1 $ & $1$ & $164$ \\ \hline
$\lbrack 030uu33u1300130u]$ & $1$ & $0$ & $158$ & $[13130u1u0u3uuuu1]$ & $1$
& $1$ & $170$ \\ \hline
$\lbrack u00uu3u033000103]$ & $1+u$ & $3$ & $176$ & $[10u000uu3031003u]$ & $1$ & $3$ & $196$ \\
\hline
\end{tabular}%
\end{center}
\end{table}

In addition to these, by applying the extension method in Theorem \ref%
{extensionA} to $\mathcal{C}_{64}$ we were able obtain 13 new codes which
are listed in Table \ref{tab:extA}.
\begin{table}[tbp]
\caption{$\left[ 68,34,12\right] $-codes in $W_{68,2}$ by Theorem \protect
\ref{extensionA} (13 codes)}
\label{tab:extA}
\begin{center}
\begin{tabular}{|l|l|l|l|}
\hline
$X$ & $c$ & $\gamma $ & $\beta $ \\ \hline
$\lbrack 30u101113u3131030u10uu0uu0111u33]$ & $1$ & $0$ & $172$ \\ \hline
$\lbrack u1uu1131u133331330uu01u1uu333100]$ & $1+u$ & $0$ & $176$ \\ \hline
$\lbrack 131333uuu0u100u101031u03313uuuu0]$ & $1$ & $1$ & $148$ \\ \hline
$\lbrack 001u0uu131u1uu3u00uuu00u0u003033]$ & $1+u$ & $1$ & $152$ \\ \hline
$\lbrack u30301u01u0u010u311303u3033u0u13]$ & $1$ & $1$ & $166$ \\ \hline
$\lbrack 033333133310u3u0u030013301131011]$ & $1$ & $1$ & $168$ \\ \hline
$\lbrack 0130330100uuu11101u3013uu111u301]$ & $1+u$ & $1$ & $172$ \\ \hline
$\lbrack 10uuuu000u30u3u3111u1uu3u3u00030]$ & $1+u$ & $1$ & $174$ \\ \hline
$\lbrack u033030311u0uuu13311uuu030uuuu01]$ & $1$ & $1$ & $176$ \\ \hline
$\lbrack uuu103031131313100033u01u3010003]$ & $1$ & $1$ & $178$ \\ \hline
$\lbrack 13uu331033u0103uuuu10uu303103133]$ & $1$ & $1$ & $190$ \\ \hline
$\lbrack 0uu0113u00111u00313u3133u1311uuu]$ & $1+u$ & $1$ & $196$ \\ \hline
$\lbrack 331u0uu3u10003uuu3u01110u0u31333]$ & $1$ & $3$ & $188$ \\ \hline
\end{tabular}%
\end{center}
\end{table}

\section{New doubly even binary codes of lengths $80$, $88$ and $96$}

In this section, specific four circulant codes over the ring $\mathbb{F}%
_{2}+u\mathbb{F}_{2}$ are considered. The codes are constructed as lifts of
binary codes. Binary images of the codes are extremal doubly-even codes of
length $80$ and $88$, the inequivalence of the codes is verified by the
invariants. As a result, we obtain \ 14 new extremal self-dual $%
[80,40,16]_{2}$ Type II codes and first extremal Type II codes of length 88
with an automorphism group of order 44. In addition, Type II $[96,48,16]_{2}$
codes with new weight enumerators are obtained by applying the method to $%
\mathbb{F}_{4}+u\mathbb{F}_{4}$.

\subsection{New extremal doubly even $\left[ 80,40,16\right] _{2}$ codes}

The weight enumerator of a doubly-even $\left[ 80,40,16\right] _{2}$ code is
uniquely determined as $1+97565y^{16}+12882688y^{20}+\cdots $ \cite%
{dougherty1}. The extended quadratic residue code $QR_{80}$ is the first
doubly-even $\left[ 80,40,16\right] _{2}$ code \cite{macwilliams}. In \cite%
{dontcheva2}, Dontcheva and Harada constructed 11 new codes with an
automorphism of order 19. Later Gulliver and Harada constructed 10 new codes
by double circulant construction in \cite{gulliver}. We construct 14 new
codes as Gray images of four circulant $\mathbb{F}_{2}+u\mathbb{F}_{2}$%
-lifts of $\left[ 40,20,8\right] _{2}$-codes given in Table \ref{tab:40codes}%
. The codes which have an automorphism group of orders $40$ and $240$ are
the first such codes. The inequivalence of the codes is checked by the
invariants. Let $c_{1},c_{2},\ldots ,c_{97565}$ be the codewords of weight $%
16$ in an extremal doubly-even $\left[ 80,40,16\right] _{2}$ code. Let $%
I_{j}=\left\vert \left\{ \left( c_{k},c_{l}\right) |\ d\left(
c_{k},c_{l}\right) =j,\ k<l\right\} \right\vert $ where $d$ is the Hamming
distance. Two codes are inequivalent if their $I_{16}$-values are different
since $I_{16}$ is invariant under a permutation of the coordinates.
\begin{table}[tbp]
\caption{$\left[ 40,20,8\right] _{2}$ four circulant self-dual codes}
\label{tab:40codes}
\begin{center}
\begin{tabular}{|l|l|l|l|l|l|}
\hline
& $r_{A}$ & $r_{B}$ & $A_{8}$ & $I_{8}$ & $\left\vert Aut\left( C\right)
\right\vert $ \\ \hline
$\mathcal{D}_{1}$ & $0100011001$ & $1110100111$ & $285$ & $2520$ & $%
2^{3}3\times 5$ \\ \hline
$\mathcal{D}_{2}$ & $0100000010$ & $1101001001$ & $285$ & $3090$ & $%
2^{3}\times 5$ \\ \hline
$\mathcal{D}_{3}$ & $0011001011$ & $0001101111$ & $285$ & $2610$ & $%
2^{2}\times 5$ \\ \hline
$\mathcal{D}_{4}$ & $1110001110$ & $0100010001$ & $285$ & $4440$ & $%
2^{14}3\times 5$ \\ \hline
$\mathcal{D}_{5}$ & $1010011111$ & $1101011100$ & $125$ & $360$ & $%
2^{3}3\times 5$ \\ \hline
$\mathcal{D}_{6}$ & $0110000110$ & $1001001110$ & $125$ & $390$ & $2^{3}5$
\\ \hline
$\mathcal{D}_{7}$ & $1100101010$ & $0110100100$ & $125$ & $570$ & $2^{2}5$
\\ \hline
\end{tabular}%
\end{center}
\end{table}

As lifts of the codes in Table \ref{tab:40codes} we obtain new extremal
doubly-even codes of length $80$ which are listed in Table \ref{tab:new80}.
\begin{table}[tbp]
\caption{New doubly even binary codes of lengths $80$ and $88$ from $\mathbb{%
F}_{2}+u\mathbb{F}_{2}$}
\label{tab:new80}
\begin{center}
\begin{tabular}{|l|l|l|l|l|l|}
\hline
$\mathcal{L}$ & $\mathcal{C}$ & $r_{A}$ & $r_{B}$ & $\left\vert Aut\left(
\mathcal{L}\right) \right\vert $ & $I_{16}$ \\ \hline
$\mathcal{L}_{80,1}$ & $\mathcal{D}_{1}$ & $\left[ 01uuu13uu3\right] $ & $%
\left[ 111010u131\right] $ & $2^{4}3\times 5$ & $20342040$ \\ \hline
$\mathcal{L}_{80,2}$ & $\mathcal{D}_{1}$ & $\left[ u3uuu130u3\right] $ & $%
\left[ 113030u133\right] $ & $2^{3}5$ & $20062500$ \\ \hline
$\mathcal{L}_{80,3}$ & $\mathcal{D}_{1}$ & $\left[ 0300033001\right] $ & $%
\left[ 13301uu313\right] $ & $2^{3}5$ & $20008440$ \\ \hline
$\mathcal{L}_{80,4}$ & $\mathcal{D}_{2}$ & $\left[ 01u0uu0u3u\right] $ & $%
\left[ 1301uu10u3\right] $ & $2^{3}5$ & $20082720$ \\ \hline
$\mathcal{L}_{80,5}$ & $\mathcal{D}_{3}$ & $\left[ 0u31u03033\right] $ & $%
\left[ uuu13u3113\right] $ & $2^{3}5$ & $20031600$ \\ \hline
$\mathcal{L}_{80,6}$ & $\mathcal{D}_{3}$ & $\left[ 0u11003033\right] $ & $%
\left[ 0uu33u1331\right] $ & $2^{4}5$ & $20195400$ \\ \hline
$\mathcal{L}_{80,7}$ & $\mathcal{D}_{4}$ & $\left[ 3330u03130\right] $ & $%
\left[ u1u003uu01\right] $ & $2^{4}5$ & $20207640$ \\ \hline
$\mathcal{L}_{80,8}$ & $\mathcal{D}_{4}$ & $\left[ 1130u03310\right] $ & $%
\left[ u3u0u10u03\right] $ & $2^{4}5$ & $20306280$ \\ \hline
$\mathcal{L}_{80,9}$ & $\mathcal{D}_{5}$ & $\left[ 1u3u011131\right] $ & $%
\left[ 31u1u113u0\right] $ & $2^{3}5$ & $20003880$ \\ \hline
$\mathcal{L}_{80,10}$ & $\mathcal{D}_{5}$ & $\left[ 3030013111\right] $ & $%
\left[ 13u3013100\right] $ & $2^{4}3\times 5$ & $20248440$ \\ \hline
$\mathcal{L}_{80,11}$ & $\mathcal{D}_{5}$ & $\left[ 1u1u011133\right] $ & $%
\left[ 33030111u0\right] $ & $2^{4}3\times 5$ & $20457960$ \\ \hline
$\mathcal{L}_{80,12}$ & $\mathcal{D}_{6}$ & $\left[ u1100u031u\right] $ & $%
\left[ 1u03u03310\right] $ & $2^{3}5$ & $19992780$ \\ \hline
$\mathcal{L}_{80,13}$ & $\mathcal{D}_{6}$ & $\left[ 0110uu031u\right] $ & $%
\left[ 3003003130\right] $ & $2^{3}5$ & $20021700$ \\ \hline
$\mathcal{L}_{80,14}$ & $\mathcal{D}_{7}$ & $\left[ 11uu303u10\right] $ & $%
\left[ 03103u010u\right] $ & $2^{3}5$ & $20043240$ \\ \hline
$\mathcal{L}_{88,1}$ & $\mathcal{C}_{88}$ & $\left[ 13303030003\right] $ & $%
\left[ 1u31u1033u3\right] $ & $2^{2}11$ & $1060092$ \\ \hline
$\mathcal{L}_{88,2}$ & $\mathcal{C}_{88}$ & $\left[ 13301u30003\right] $ & $%
\left[ 303103u33u1\right] $ & $2^{2}11$ & $1078803$ \\ \hline
$\mathcal{L}_{88,3}$ & $\mathcal{C}_{88}$ & $\left[ 11303u3u001\right] $ & $%
\left[ 1u3101u13u1\right] $ & $2^{2}11$ & $1089990$ \\ \hline
$\mathcal{L}_{88,4}$ & $\mathcal{C}_{88}$ & $\left[ 331u1u1u0u3\right] $ & $%
\left[ 3u31u3u3303\right] $ & $2^{2}11$ & $1095666$ \\ \hline
$\mathcal{L}_{88,5}$ & $\mathcal{C}_{88}$ & $\left[ 13101u3uu01\right] $ & $%
\left[ 3u33u103103\right] $ & $2^{2}11$ & $1103553$ \\ \hline
$\mathcal{L}_{88,6}$ & $\mathcal{C}_{88}$ & $\left[ 311u3010u03\right] $ & $%
\left[ 1011u3u31u1\right] $ & $2^{2}11$ & $1115400$ \\ \hline
$\mathcal{L}_{88,7}$ & $\mathcal{C}_{88}$ & $\left[ 311010300u1\right] $ & $%
\left[ 3u33u1u31u3\right] $ & $2^{2}11$ & $1132164$ \\ \hline
$\mathcal{L}_{88,8}$ & $\mathcal{C}_{88}$ & $\left[ 33103u30001\right] $ & $%
\left[ 1u11u301303\right] $ & $2^{2}11$ & $1115664$ \\ \hline
$\mathcal{L}_{88,9}$ & $\mathcal{C}_{88}$ & $\left[ 333u1u3uu03\right] $ & $%
\left[ 101103031u1\right] $ & $2^{2}11$ & $1128402$ \\ \hline
$\mathcal{L}_{88,10}$ & $\mathcal{C}_{88}$ & $\left[ 31103u1uu01\right] $ & $%
\left[ 3u3101033u3\right] $ & $2^{2}11$ & $1160181$ \\ \hline
\end{tabular}%
\end{center}
\end{table}
The codes $\mathcal{L}_{80,6},\mathcal{L}_{80,7}$ and $\mathcal{L}_{80,8}$
in Table \ref{tab:new80} have an automorphism group of order $80$ and these
are inequivalent to such codes $P_{80,2},P_{80,3},P_{80,4}$ and $P_{80,5}$
in \cite{gulliver} since their $I_{16}$ values are $20290440$, $20187210$, $%
20201130$ and $20034000$ respectively. Hence, we have the following theorem:

\begin{theorem}
There exist at least $36$ extremal doubly-even self-dual codes of length $80$%
.
\end{theorem}

By the Assmus-Matson theorem the codewords of weight $16$ in an extremal
doubly-even code of length $80$ form a $3$-design.

\begin{theorem}
There are at least $36$ non-isomorphic $3-\left( 80,16,665\right) $ desings.
\end{theorem}

\subsection{New extremal doubly even $\left[ 88,44,16\right] _{2}$ codes}

There are four circulant $\left[ 44,22,8\right] _{2}$ self-dual codes. We
apply the lifting method to one of them and obtain $100$ inequivalent
extremal doubly even $\left[ 88,44,16\right] _{2}$ codes as the Gray images
of $\mathbb{F}_{2}+u\mathbb{F}_{2}$-lifts. Let $\mathcal{C}_{88}$ be the
binary four circulant code with $r_{A}=\left( 11101010001\right) $ and $%
r_{B}=\left( 10110101101\right) $. $\mathcal{C}_{88}$ is lifted to $\mathbb{F%
}_{2}+u\mathbb{F}_{2}$ and $\left[ 88,44,16\right] _{2}$ extremal doubly
even codes with an automorphism group of order $44$ are obtained. The codes
are different than the previous codes since the order of the automorphism
group is different. The inequivalence of the codes is verified by the
invariants $I_{16}$ as was done previously. In order to save space we list $%
10$ of the codes in Table \ref{tab:new80}.

\subsection{New doubly even $\left[ 96,48,16\right] _{2}$ codes}

A self-dual doubly even $\left[ 96,48,16\right] _{2}$-code has weight
enumerator $1+\left( -28086+\alpha \right) y^{16}+\left( 3666432-16\alpha
\right) y^{20}+\cdots .$ The first such code with $\alpha =37722$ is
constructed in \cite{feit} by a construction from extended binary quadratic
residue codes of length $32$ and $25$ new codes are constructed in \cite%
{dontcheva} via automorphisms of order $23$. Let $\mathcal{C}_{96}$ be the
four circulant code over $\mathbb{F}_{4}$ with $r_{A}=\left( \omega
,1,0,1+\omega ,\omega ,\omega \right) $ and $r_{B}=\left( 0,1+\omega
,1,0,1+\omega ,0\right) $. $\mathcal{C}_{96}$ is a self-dual code which has
binary Gray image a $\left[ 48,24,8\right] _{2}$. By lifting this code to $%
\mathbb{F}_{4}+u\mathbb{F}_{4}$ a family of self-dual codes obtained. As
binary Gray images of these codes a substantial number of new doubly-even
self dual $[96,48,16]_{2}$ codes\ are obtained, in order to save space just
ten of them are listed in Table \ref{tab:new96}.
\begin{table}[tbp]
\caption{New $\left[ 96,48,16\right] _{2}$ doubly even codes from $\mathbb{F}%
_{4}+u\mathbb{F}_{4}$}
\label{tab:new96}
\begin{center}
\begin{tabular}{|l|l|l|l|}
\hline
$\mathcal{L}_{96,i}$ & $r_{A}$ & $r_{B}$ & $\alpha $ \\ \hline
$\mathcal{L}_{96,1}$ & $\left( b_{3},a_{1},z_{1},c_{4},b_{4},b_{2}\right) $
& $\left( z_{1},c_{2},a_{2},z_{4},c_{4},z_{1}\right) $ & 36864 \\ \hline
$\mathcal{L}_{96,2}$ & $\left( b_{1},a_{4},z_{1},c_{1},b_{1},b_{1}\right) $
& $\left( z_{2},c_{4},a_{1},z_{2},c_{3},z_{3}\right) $ & 36876 \\ \hline
$\mathcal{L}_{96,3}$ & $\left( b_{2},a_{4},z_{1},c_{3},b_{2},b_{3}\right) $
& $\left( z_{3},c_{4},a_{4},z_{2},c_{2},z_{4}\right) $ & 36888 \\ \hline
$\mathcal{L}_{96,4}$ & $\left( b_{4},a_{2},z_{4},c_{3},b_{3},b_{3}\right) $
& $\left( z_{3},c_{1},a_{3},z_{4},c_{1},z_{3}\right) $ & 36900 \\ \hline
$\mathcal{L}_{96,5}$ & $\left( b_{1},a_{4},z_{2},c_{1},b_{3},b_{3}\right) $
& $\left( z_{3},c_{1},a_{3},z_{2},c_{2},z_{2}\right) $ & 36912 \\ \hline
$\mathcal{L}_{96,6}$ & $\left( b_{4},a_{1},z_{3},c_{2},b_{4},b_{1}\right) $
& $\left( z_{4},c_{2},a_{3},z_{3},c_{4},z_{3}\right) $ & 36936 \\ \hline
$\mathcal{L}_{96,7}$ & $\left( b_{2},a_{1},z_{2},c_{1},b_{2},b_{4}\right) $
& $\left( z_{1},c_{1},a_{2},z_{2},c_{1},z_{3}\right) $ & 36948 \\ \hline
$\mathcal{L}_{96,8}$ & $\left( b_{1},a_{1},z_{1},c_{1},b_{3},b_{1}\right) $
& $\left( z_{2},c_{3},a_{1},z_{3},c_{1},z_{1}\right) $ & 36960 \\ \hline
$\mathcal{L}_{96,9}$ & $\left( b_{2},a_{3},z_{4},c_{2},b_{4},b_{2}\right) $
& $\left( z_{2},c_{2},a_{4},z_{1},c_{1},z_{1}\right) $ & 36972 \\ \hline
$\mathcal{L}_{96,10}$ & $\left( b_{4},a_{1},z_{3},c_{4},b_{3},b_{2}\right) $
& $\left( z_{2},c_{1},a_{4},z_{1},c_{2},z_{2}\right) $ & 36984 \\ \hline
\end{tabular}%
\end{center}
\end{table}
The codes in the Tables \ref{tab:new80} and \ref{tab:new96} are generated by
$\left[ \enspace I_{2n}\enspace%
\begin{array}{|cc}
A & B \\
B^{T} & A^{T}%
\end{array}%
\right] $ over $\mathbb{F}_{2}+u\mathbb{F}_{2}$ and $\mathbb{F}_{4}+u\mathbb{%
F}_{4}$ respectively.

\section{Conclusion}
The binary extension theorems in the literature are used to obtain binary self-dual codes of length $n+2$ from self-dual codes of length $n$. They have been effectively used to characterize many extremal binary self-dual codes.

In our work, we generalized this extension to rings, since recently self-dual codes over rings have been used to obtain extremal binary self-dual codes. The extension theorems that we suggest can be applied to all rings of characteristic 2. By using a family of such rings, i.e., $\F_{2^m}+u\F_{2^m}$, with $m=1,2$ and the aforementioned extension theorems we were able to obtain a substantial number of new binary extremal self-dual codes of certain lengths, the results of which have been tabulated throughout the paper. The results indicate the effectiveness of these extension theorems and thus we believe it will add to the motivation of studying self-dual codes over rings. Working out these extensions in different rings might fill out a lot of the gaps in the study of extremal binary self-dual codes.

A possible line of research could be attempting such extension theorems for rings of other characteristic as well, such as $\mathbb{Z}_4$.

\end{document}